\documentclass[a4paper,11pt]{article}
\usepackage{pos}
\usepackage{placeins}
\providecommand{\cov}{\operatorname{Cov}}

\title{The NRQCD $\Upsilon$ spectrum at non-zero temperature using Backus-Gilbert regularisations}

\author*[a]{Antonio Smecca}
\author[a]{Gert Aarts}
\author[a]{Chris Allton}
\author[a,b]{Ryan Bignell}
\author[a]{Timothy J.~Burns}
\author[c]{Benjamin J\"ager}
\author[d]{Rachel Horohan D’Arcy}
\author[e]{Seyong Kim}
\author[f]{Maria-Paola Lombardo}
\author[a]{Ben Page}
\author[b]{Sin\'ead M.~Ryan}
\author[a]{Tom Spriggs}
\author[b,d]{Jon-Ivar Skullerud}

\affiliation[a]{Department of Physics, Swansea University, Swansea, SA2 8PP, United Kingdom }

\affiliation[b]{School of Mathematics, Trinity College, Dublin, Ireland}
\affiliation[c]{Quantum Field Theory Center \& Danish IAS, Department of Mathematics and
Computer Science , University of Southern Denmark, 5230, Odense M, Denmark}
\affiliation[d]{Department of Physics and Hamilton Institute, National University
  of Ireland Maynooth, County Kildare, Ireland}
\affiliation[e]{Department of Physics, Sejong University, Seoul 143-747, Korea}
\affiliation[f]{INFN, Sezione di Firenze, 50019 Sesto Fiorentino (FI), Italy}

\emailAdd{antonio.smecca@swansea.ac.uk}

\abstract{Understanding how the properties of heavy mesons change as temperature increases is crucial for gaining valuable insights into the quark-gluon plasma. Information about meson masses and decay widths is encoded in the meson spectral function, which, in principle, can be extracted from Euclidean correlation functions via generalised Laplace transformations. However, this inverse problem is ill-posed for lattice correlation functions and requires regularisation. In this work, we present the latest results for bottomonium spectral functions obtained within the lattice NRQCD framework using the Backus-Gilbert regularisation, along with two other variants, one of which is commonly referred to as the HLT method. Our analysis employs Generation 2L anisotropic lattice configurations produced by the \textsc{Fastsum} collaboration.}

\FullConference{The 41st International Symposium on Lattice Field Theory (LATTICE2024)\\
 28 July - 3 August 2024\\
Liverpool, UK\\}

\begin{document}
\maketitle

\section{Introduction}
Understanding the properties of heavy mesons at high temperatures is very important to gain insights on the QCD phase diagram. In particular, the dissociation of heavy quarkonia in a deconfined medium may serve as a thermometer for relativistic heavy-ion collisions~\cite{Matsui:1986dk,Rothkopf:2019ipj}.
The abundant production of bottomonium systems in LHC heavy-ion collision experiments make them good probes of the quark-gluon plasma (QGP). Results from the CMS Collaboration indicate sequential suppression in this system~\cite{CMS:2011all}.

For more than a decade, the \textsc{Fastsum} collaboration has extensively studied bottomonium systems using lattice QCD simulations with anisotropic lattices and Wilson fermions~\cite{Aarts:2010ek,Aarts:2011sm,Aarts:2014cda,Aarts:2013kaa}.
Of particular interest in this investigation is the spectral function $\rho(\omega)$, which encodes information about meson masses and decay widths.
This function can be extracted from Euclidean correlation functions through an inverse Laplace transform. However, the inverse problem is ill-posed when using lattice correlation functions, and several different methods have been developed over the years to overcome this obstacle.

This work builds on previous studies~\cite{Page:2022tbj,Skullerud:2022yjr,FASTSUM:2022qzx,Spriggs:2021dsb,Page:2021ohe} of the bottomonium spectral functions using linear approaches to the inverse problem based on the Tikhonov and Backus-Gilbert methods~\cite{backusResolvingPowerGross1968,gilbertUniquenessInversionInaccurate1970,tikhonovStabilityInverseProblems1943}. For other studies of bottomonium spectral functions by the \textsc{Fastsum} collaboration with non-linear methods see Refs.~\cite{Spriggs:2021jsh,Offler:2021fmg}.

\section{Lattice ensembles}
For this investigation, we used the Generation 2L gauge ensembles generated by the \textsc{Fastsum} collaboration.
The Wilson-clover fermion action was employed, with tree-level tadpole improvement with stout links, same parameters as the HadSpec collaboration~\cite{Edwards:2008ja}, approximately 1000 configurations for each temperature, $m_{\pi} = 239(1)$ $\mathrm{MeV}$, anisotropic lattices with anisotropy $\xi = 3.453(6)$ and $T_{c} = 167(2)$ $\mathrm{MeV}$. In this work, we use the (earlier) scale setting from Ref.~\cite{Wilson:2015dqa}. As such, the spatial lattice spacing is $a_s = 0.1136(6)$ fm.
The strange quark has been tuned to its physical value via the tuning of the light and strange pseudoscalar masses~\cite{HadronSpectrum:2012gic,Cheung:2016bym}. The estimate for $T_{\rm pc}$ comes from an analysis of the renormalised chiral condensate, which uses the updated lattice spacing of Ref.~\cite{Aarts:2020vyb}, that was implemented in our analysis in Ref.~\cite{,Aarts:2022krz}.

The bottom quarks are simulated using the non-relativistic QCD (NRQCD) effective theory~\cite{Lepage:1992tx} with power counting in the heavy quark velocity in the bottomonium rest frame, $v \sim |\boldsymbol{p}|/m_b$.
In the NRQCD formalism only energy differences are physically significant, such that we can define the additive mass renormalisation from the following mass difference, finding
\begin{align}
  E_0 = M_{\rm exp}(\Upsilon) - M(\Upsilon) = 7463~\mathrm{MeV},
\end{align}
where the lattice value of $M(\Upsilon)$ was determined using a single exponential fit to the zero-temperature correlator in the vector channel.

\begin{table}[t]
  \centering
  \begin{tabular}{r|rrrrr||rrrrrrr}
    $N_\tau$ & 128 & 64 & 56 & 48 & 40 & 36 & 32 & 28 & 24 & 20 & 16 & 12\\ \hline
    $T (\mathrm{MeV})$ & 47 & 95 & 109 & 127 & 152 & 169 & 190 & 217 & 253 & 304 & 380 & 507\\
  \end{tabular}
  \caption{\textsc{Fastsum} Generation 2L ensembles used in this work. The lattice size is $32^3 \times N_\tau$, with temperature $T = 1/(a_\tau N_\tau)$. The temporal and spatial lattice spacing are respectively $0.03246(7)$ and $a_s = 0.11208(31)$ fm giving a renormalised anisotropy $\xi = a_s/a_\tau = 3.453(6)$ and the pion mass is $m_\pi = 239(1)$ MeV~\cite{Wilson:2019wfr}. The estimate for $T_{\rm pc}$ comes from an analysis of the renormalised chiral condensate and equals $T_{\rm pc} = 167(2)(1)$ MeV~\cite{Aarts:2020vyb,Aarts:2022krz}. Full details of these ensembles may be found in Refs.~\cite{Aarts:2020vyb,Aarts:2022krz,Gen2Lzenodo}.
  }
  \label{tab:ensembles}
\end{table}

\section{NRQCD spectral functions}
To study the spectrum of bottomonium at high temperatures the best way forward is to employ methods that do not make any assumption on the spectral structure. For this reason, the observable of interest here is the spectral function $\rho(\omega)$. Extracting $\rho(\omega)$ from lattice correlators is an ill-posed problem that has recently attracted a lot of attention within the lattice community, as evidenced by three plenary presentations on the topic~\cite{Jay:2025dzl,Liang:2020sqi,Bulava:2023mjc}. The access to $\rho(\omega)$ unlocked several important investigations using lattice simulations, such as the $R$-ratio for $e^+ e^- \rightarrow$ hadrons~\cite{ExtendedTwistedMassCollaborationETMC:2022sta}, inclusive hadronic decays of the $\tau$ lepton~\cite{ExtendedTwistedMass:2024myu,Evangelista:2023fmt} and inclusive semi-leptonic decays of heavy mesons~\cite{Gambino:2022dvu,Kellermann:2024jqg,Barone:2023iat,Kellermann:2023yec,Barone:2023tbl,Gambino:2020crt}. Other applications include neutrino-nucleon scattering~\cite{Fukaya:2020wpp,Liang:2023uai,Liang:2019frk} and advances in hadron spectroscopy~\cite{Panero:2023zdr,Forzano:2024kjh,Bennett:2024cqv}. For a study of the inverse problem in the continuum limit see Ref.~\cite{Bruno:2024fqc}.

In the relativistic formulation of lattice QCD, one can relate the Euclidean correlator to the spectral function through the following integral equation
  
\begin{align}
  G(\tau) = \int_{0}^{\infty} \frac{d\omega}{2\pi}~\rho(\omega)~K(\omega, \tau, T),
  \label{eq:Relativistic}
\end{align}
where $G(\tau)$ is the meson correlation function, and $K(\omega, \tau, T)$ is an integration kernel that depends on the temperature of the system.

In the NRQCD formalism one can re-write equation~\ref{eq:Relativistic} as
\begin{align}
  G(\tau) = \int_{\omega_{\min}}^{\omega_{\max}} d\omega~\rho(\omega)~e^{-\omega \tau},
  \label{eq:NRQCD}
\end{align}
where $\omega_{\max}$ is the energy cutoff and $\omega_{\min}$ is the minimum energy chosen to perform the integration~\cite{Burnier:2007qm,Aarts:2010ek,Aarts:2011sm}.

Due to the finite volume of lattice simulations, the spectral density contained in LQCD correlators is a distribution that in order to be physically relevant needs to be convoluted with a \textit{smeared} kernel. Smearing is needed to turn the distribution of $\delta$-functions into a smooth function that can be studied in the infinite-volume limit~\cite{Hansen:2017mnd}. Assuming an infinite number of time-slices $\tau$, the Stone-Weierstrass theorem admits an exact polynomial expression for these smearing kernels
\begin{align}
  \overline{\Delta}(\omega,\omega_n) = \sum_{\tau=1}^{\infty} g_{\tau}(\omega_n)e^{-\omega \tau} \approx \delta(\omega-\omega_n),
  \label{eq:Kernel}
\end{align}
which gives the \underline{exact} model independent solution
\begin{align}
  \rho(\omega_n) = \int_{\omega_{\min}}^{\infty} d\omega~\delta(\omega-\omega_n)\rho(\omega).
\end{align}
This is the central idea behind any linear method for spectral reconstruction.

Imagining to be in an \textit{ideal} scenario where we have access to the exact correlator at infinite discrete times, one can obtain the spectral function following
\begin{align}
  \rho(\omega_n) &= \int_{\omega_{\min}}^{\infty} d\omega~\overline{\Delta}(\omega,\omega_n)\rho(\omega)
  = \int_{\omega_{\min}}^{\infty} d\omega~\sum_{\tau=1}^{\infty}g_{\tau}(\omega_n)e^{-\omega \tau}\rho(\omega)
  = \sum_{\tau=1}^{\infty}g_{\tau}(\omega_n)G(\tau),
\end{align}
where in the last step we used the definition for $G(\tau)$ in equation~\ref{eq:NRQCD}.

\subsection{Backus-Gilbert and Tikhonov methods}
In realistic lattice QCD simulations, meson correlation functions are truncated by the maximum time extent of the lattice and are affected by statistical noise. This means that equation~\ref{eq:Kernel} can only be approximated with the given data. The consequence of this is that we can at most get an estimate for the spectral function as
\begin{align}
  \hat{\rho}(\omega_n) = \sum_{\tau=1}^{\tau_{\max}}g_{\tau}(\omega_n)G(\tau),
\end{align}
where $\tau_{\max}$ is the maximum time extent of the lattice and $\hat{\rho}(\omega_n)$ represents the estimate of $\rho(\omega)$ at $\omega_n$.

The essential ingredient for the reconstruction of the spectral function is the calculation of the coefficients $g_{\tau}(\omega_n)$. Following standard methods like the Backus-Gilbert~\cite{backusResolvingPowerGross1968,gilbertUniquenessInversionInaccurate1970} or Tikhonov~\cite{tikhonovStabilityInverseProblems1943}, they are obtained minimising the functional

\begin{align}
  A^{\rm BG,\rm Tikh}[\boldsymbol{g}] = \int_{\omega_{\min}}^{\omega_{\max}} d\omega~[\overline{\Delta}^{\rm BG,\rm Tikh}(\omega,\boldsymbol{g})-\delta(\omega-\omega_n)]^2,
\end{align}
which can be understood as finding the smearing kernel closest to a $\delta$-function. In the equation above, $\boldsymbol{g}$ corresponds to a vector of coefficients at different $\tau$ for a given $\omega$. Our results were obtained setting $a_{\tau}\omega_{\min}=-0.1$ and $a_{\tau}\omega_{\max}=1.0$.
To preserve the scale of the spectrum, the chosen representation for the width functional is typically minimised in conjunction with a unit area constraint.
In this case, the kernel approximation uses Legendre polynomials. There is an interesting alternative method based on Chebyshev polynomials~\cite{Bailas:2020qmv} which we will not be discussed here.

The fact that we are facing an ill-posed (or rather ill-conditioned) problem is that the excessive minimisation of $A[\boldsymbol{g}]$ leads to gigantic coefficients, which once applied to the meson correlator lead to huge statistical errors, making the result useless from a physical point of view. An example of this can be seen in Figure~2 of Ref.~\cite{Hansen:2019idp}.
To prevent the coefficients $g_{\tau}$ from reaching such large values, one needs to introduce an error functional $B[\boldsymbol{g}]$ and minimise the linear combination
\begin{align}
  W[\boldsymbol{g}] = \frac{A[\boldsymbol{g}]}{A[\boldsymbol{0}]} + \lambda B[\boldsymbol{g}],
\end{align}
where the error functional is defined as
\begin{align}
  B^{\rm BG}[\boldsymbol{g}] = \frac{\cov_{\tau,\tau'}[\boldsymbol{g}]}{G(\tau=0)^2}, \hspace{1.5cm} B^{\rm Tikh} [\boldsymbol{g}] = \mathbb{I},
\end{align}
for the Backus-Gilbert and Tikhonov methods respectively, where $\cov$ is the covariance matrix of the correlator and $\mathbb{I}$ is the identity matrix. The hyper-parameter $\lambda$ determines how much the minimisation is being regularised. An excessive regularisation introduces large systematic errors due to a poor spectral function reconstruction while no regularisation leads to gigantic coefficients. Therefore, it is essential to find the optimal trade-off value for $\lambda$.

\subsection{HLT method}
The HLT method~\cite{Hansen:2019idp} is a modification of the Backus-Gilbert method. In Refs.~\cite{DelDebbio:2024sfa,DelDebbio:2024lwm,Lupo:2023qna} the authors show an important equivalence between this method and the approach based on Gaussian processes. The error functional is indeed the same as $B^{\rm BG}[\boldsymbol{g}]$, while the $A[\boldsymbol{g}]$ reads
\begin{align}
  A^{\rm HLT}_n[g_{\tau}] = \int_{\omega_{\min}}^{\infty}d\omega~e^{\alpha \omega}\left|\overline{\Delta}_{\sigma}^{\rm HLT} - \Delta^{\rm HLT}_{\sigma}\right|^2,
\end{align}
where the $\delta$-function has been replaced by $\Delta^{\rm HLT}_{\sigma}$ and $\alpha$ is a hyper-parameter selecting the type of Jacobi polynomial used in the approximation. In this work we used $\alpha=\{0, 1, -1.99\}$ to select different polynomials and change the speed of convergence of the approximation. The kernel $\Delta^{\rm HLT}_{\sigma}$ is a target kernel which is chosen \textit{a priori} to be a smeared version of the $\delta$-function with a smearing radius $\sigma$ which enters as an input. The minimisation then gives the best approximation of the smearing kernel $\Delta^{\rm HLT}_{\sigma}$.
Results are obtained using $a_{\tau}\omega_{\min} \approx 0.05$ and $a_{\tau}\omega_{\max}=\infty$, however, little to no change is observed changing $\omega_{\min}$.

\section{Results}
\begin{figure}[ht]
  \centering
  \includegraphics[width=0.49\textwidth]{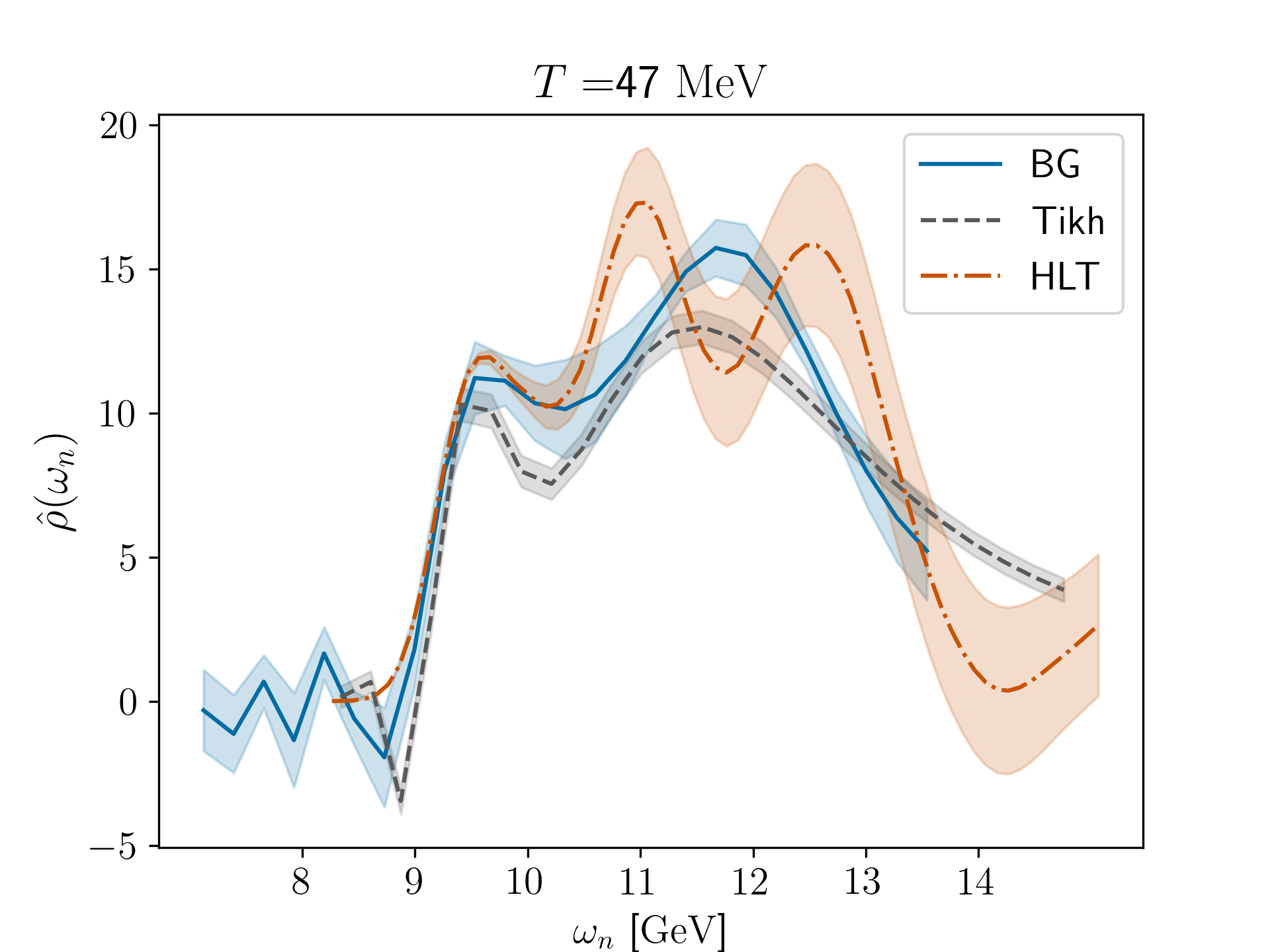}
  \includegraphics[width=0.49\textwidth]{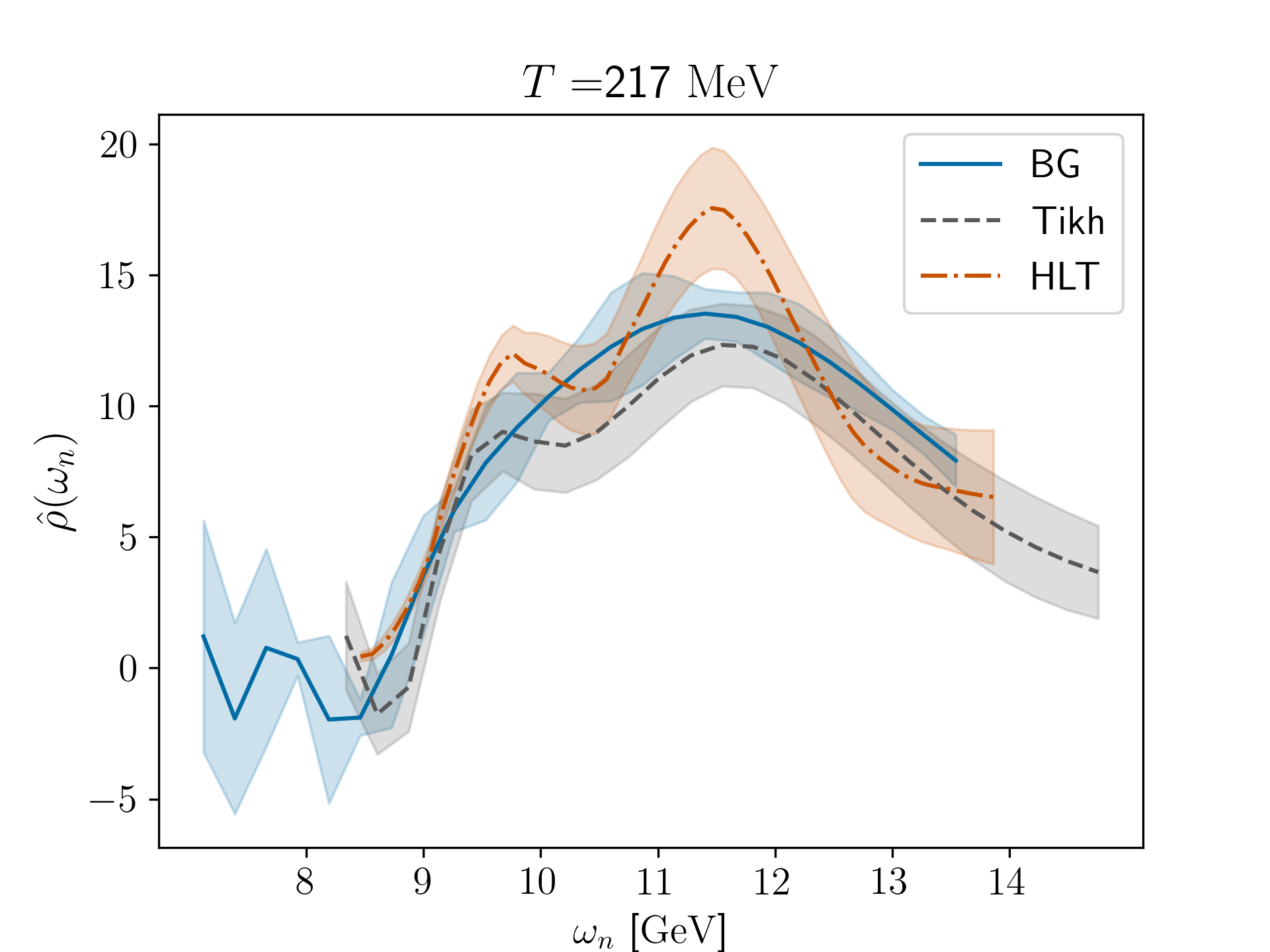}
  \caption{$\Upsilon$ spectral function using the Backus-Gilbert (blue solid line), Tikhonov (dark-gray dashed line) and HLT (dark-orange dashed-dotted line) methods at different values of the temperature.}
  \label{fig:method}
\end{figure}

All the linear methods presented above should in principle agree within the given uncertainties, especially for the first peak of the spectral function, corresponding to the ground state. The agreement is found for both low and high temperatures, as shown in Figure~\ref{fig:method}.

The results corresponding to the Tikhonov method are obtained after applying the Laplace shift technique introduced in Ref.~\cite{Page:2021ohe}. While the Laplace shift has little to no effect in the Backus-Gilbert and HLT results, it was found empirically to be beneficial for the Tikhonov regularisation. Specifically, the Laplace shift facilitates a faster scan of the hyper-parameter $\lambda$. Indeed, one can show that the application of the Laplace shift is equivalent to a change in the coefficients $g_{\tau}$ due to a change in $\lambda$.

Information about the ground state mass and decay width of $\Upsilon$ can be extracted by fitting the spectral function to a model Gaussian function. For the Backus-Gilbert and Tikhonov results, the fit is done using all the points leading to the first peak of the spectral function. For the HLT method, we adopt the procedure introduced in Ref.~\cite{DelDebbio:2022qgu}, which exploits the fact that we know \textit{a priori} the width of the Gaussian function used to smear the spectral function. This implies that we cannot obtain the decay width from the fit. By checking the quality of the HLT spectral reconstruction we try to use the smallest possible value of $\sigma$ allowed by the data.

All linear approaches to the inverse problem rely exclusively on the information provided by the given data points. As a result, the quality of the reconstruction progressively deteriorates as the number of time-slices is reduced. In our fixed scale setup, where the temperature of the lattice simulation is increased by reducing the temporal lattice extent, this implies that the quality of the spectral function reconstruction progressively worsens with increasing temperature.
Unfortunately, this hinders any conclusive statements about the behaviour of spectral functions at high temperatures, which is the region we are most interested in.
An imperfect reconstruction inevitably affects the accurate determination of both the position and width of the peaks in the spectral function. Therefore, it is unclear if the broadening of the spectral features is a thermal effect or an artifact of the reconstruction process. In the HLT method, the progressive degradation of the reconstruction forces the use of larger smearing parameter values.

\begin{figure}
  \centering
  \includegraphics[width=.49\textwidth]{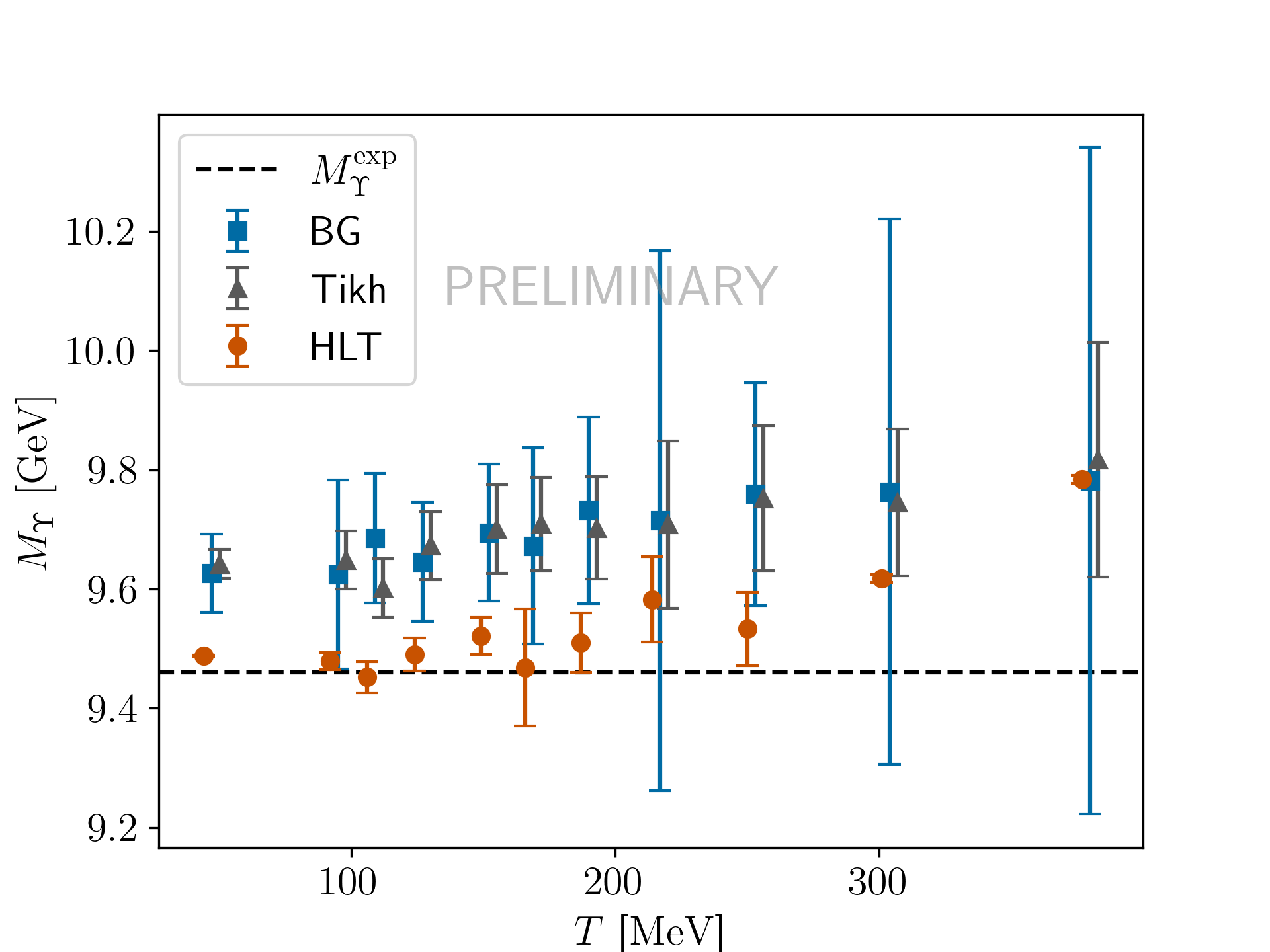}
  \includegraphics[width=.49\textwidth]{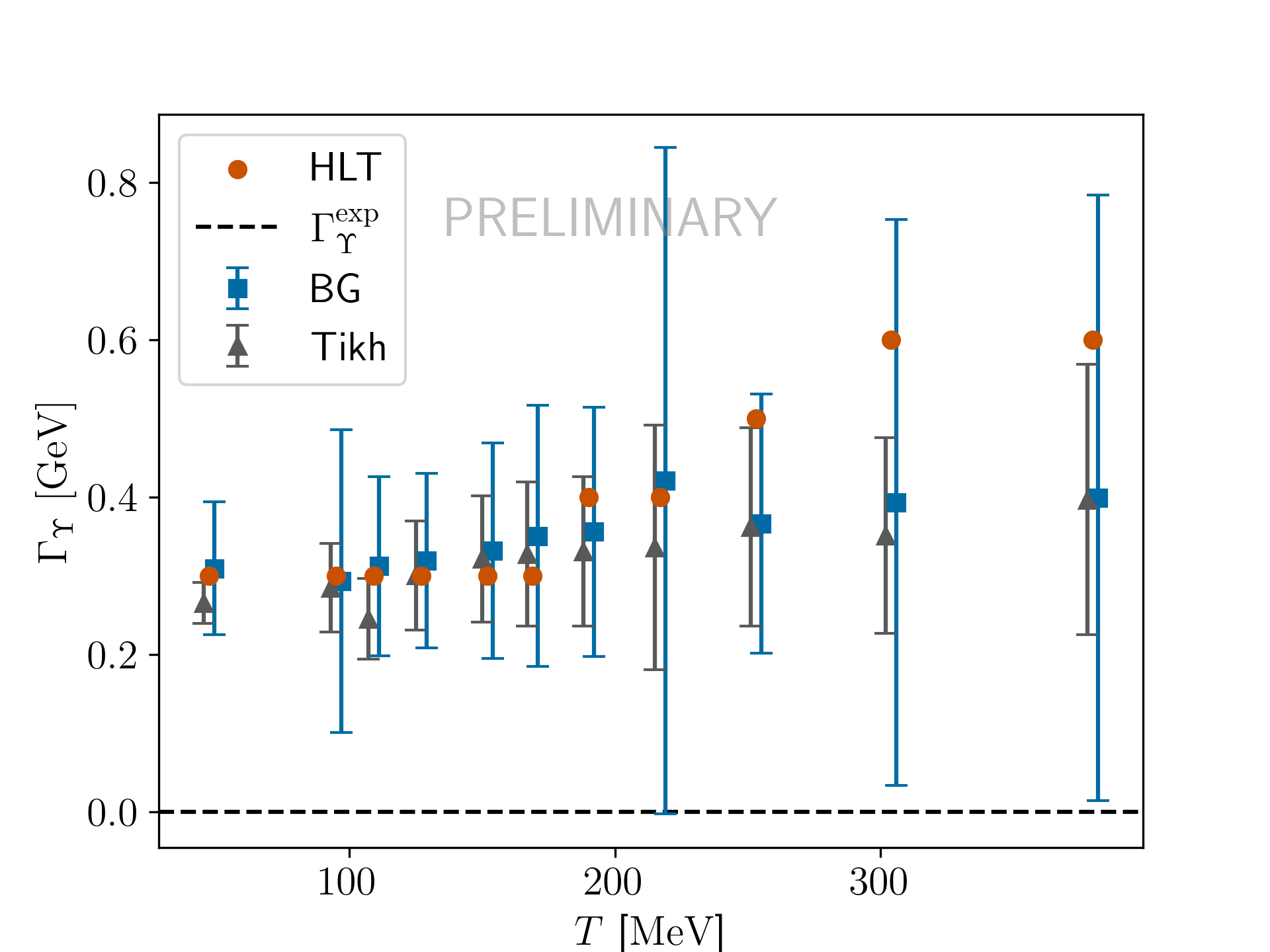}
  \caption{Results of the ground state mass and decay width using the Backus-Gilbert (blue squares), Tikhonov (dark-gray triangles) and HLT (dark-orange circles) methods as a function of temperature. HLT results for the width do not show errorbars since they correspond to input parameters in the minimisation procedure. Results are shifted horizontally to improve readability. The black dashed line correspond to the experimental values at zero temperature.}
  \label{fig:final_res}
\end{figure}

Figure~\ref{fig:final_res} shows the results of the ground state mass and decay width obtained by each method considered in this work. We stress once again that the HLT values of the width are not physical predictions but input parameters.


\section{Discussion and Conclusion}
The values of the ground state mass show excellent agreement between the Backus-Gilbert and Tikhonov methods and are compatible with the results of the HLT method only for high enough temperatures. At lower temperatures, the HLT method results are more consistent with the experimental value at zero temperature. Overall, the HLT results show a better precision compared to the values obtained with the Backus-Gilbert and Tikhonov methods.
This discrepancy can be attributed to the different fitting procedures employed. The lack of a clear peak structure in the spectral function makes the fits of Backus-Gilbert and Tikhonov more susceptible to unwanted excited states which can alter the accurate and precise determination of the ground state peak. This issue is less severe for HLT fits which benefit from the additional information on the smearing width to better constrain the energy states in the spectral function.

The only physical prediction for the decay widths come from the Backus-Gilbert and Tikhonov methods, since, in the HLT method, the width coincides with the smearing width used as input parameter in the regularisation of the inverse problem. Remarkably, the decay widths obtained with all three methods are largely compatible.
This suggests that the increase of $\sigma$ in the HLT method is well justified, considering that the two other methods, which do not rely on fixing the smearing width, obtain similar results.

Unlike the ground state mass, the numerical results for the decay width are significantly larger than the experimental result at zero-temperature, even for the coldest ensemble where thermal effects are negligible. This discrepancy arises due to the finite amount of data available giving a good approximation but never the exact spectral function.
The approximation, and consequently the width of the smearing function, depends strongly on the quality of the data. With current state-of-the-art correlators the narrowest smearing function possible has a width corresponding to $\approx 0.3$ $\mathrm{GeV}$, which is several orders of magnitude larger than the physical decay width of bottomonium.

The linear methods discussed here provide a completely model-independent approach to the inverse problem. However, they often yield a more conservative estimate of the spectral function compared to non-linear approaches. In other words, spectral functions obtained with non-linear approaches exhibit a clearer and more pronounced peak structure compared to the results of linear methods. Furthermore, the worsening of the spectral reconstruction as we increase the temperature of the system makes the unambiguous identification of thermal effects particularly challenging.

The results of this work will contribute to a comparative study of spectral reconstruction methods to spectral functions at high temperatures, currently being prepared by the \textsc{Fastsum} collaboration.

\begin{acknowledgments}
  We are grateful to the HadSpec collaboration for the use of their zero temperature ensembles. A.~S. is grateful to Marco Panero, Nazario Tantalo, Alessandro Lupo, Alessandro De Santis, Alessandro Barone and Niccol\`o Forzano for the numerous stimulating conversations on spectral reconstruction techniques.
  This work is supported by the UKRI Science and Technology Facilities Council (STFC) Consolidated Grant No. ST/X000648/1.
\end{acknowledgments}

\bibliography{references}
\bibliographystyle{JHEP}

\end{document}